\documentstyle[aps,prl,floats]{revtex}

\def\b{\beta}

\def\d{\delta}

\def\L{\Lambda}
\def\m{\mu}
\def\n{\nu}

\def\th{\theta}

\def\beq{\begin{equation}}\def\eeq{\end{equation}}
\def\beqa{\begin{eqnarray}}\def\eeqa{\end{eqnarray}}
\def\barr{\begin{array}}\def\earr{\end{array}}

\def\del{\partial}

\def\Ds {{D \hspace{-6.4pt} \slash}\;}

 \let\br=\bigr
 
\let\bm=\bibitem

\def\nn{\nonumber}
\def\bd{\begin{document}}
\def\ed{\end{document}}
\def\ba{\begin{array}}
\def\ea{\end{array}}
\def\bea{\begin{eqnarray}}
\def\eea{\end{eqnarray}}
\def\ft#1#2{{\textstyle{{\scriptstyle #1}\over {\scriptstyle #2}}}}
\def\fft#1#2{{#1 \over #2}}
\newcommand{\be}{\begin{equation}}
\newcommand{\ee}{\end{equation}}
\newcommand{\eq}[1]{(\ref{#1})}
\def\eqs#1#2{(\ref{#1}-\ref{#2})}
\def\det{{\rm det\,}}
\def\tr{{\rm tr}}
\newcommand{\ho}[1]{$\, ^{#1}$}
\newcommand{\hoch}[1]{$\, ^{#1}$}
\def\ra{\rightarrow}

\def\Xh{\hat{X}}
\def\ah{\hat{a}}
\def\xh{\hat{x}}
\def\yh{\hat{y}}
\def\ph{\hat{p}}
\def\G{{\cal G}}
\def\Dth{{\Delta_\th}}

\def\bk{{\bf k}}
\def\bx{{\bf x}}
\def\br{{\bf r}}

\begin{document}

\twocolumn[\hsize\textwidth\columnwidth\hsize\csname@twocolumnfalse\endcsname

\title{Remarks on Inflation and Noncommutative Geometry}

\author{Chong-Sun Chu $^1$, Brian R. Greene $^2$,
Gary Shiu $^3$}
\address{$^1$ Institute of Physics, University of Neuch\^atel,
CH-2000 Neuch\^atel, Switzerland}
\address{$^1$ Center for Particle Theory,
Department of Mathematical Sciences, University of Durham, DH1 3LE, UK}
\address{$^2$ Institute for Strings, Cosmology and Astroparticle
Physics, Columbia University,
New York, NY 10027, USA}
\address{$^3$ Department of Physics and Astronomy, University of
Pennsylvania,
Philadelphia, PA 19104, USA
\\Email:     chong-sun.chu@iph.unine.ch$^1$,
greene@phys.columbia.edu$^2$,
shiu@dept.physics.upenn.edu$^3$}

\maketitle

{\tighten
\begin{abstract}
We briefly discuss some possible cosmological implications of
noncommutative geometry.
While the noncommutativity we consider
does not affect gravity, it can play an important
role in the dynamics of other fields that are present
in the early universe.
We point out the possibility
that noncommutativity may cause inflation induced
fluctuations to become non-gaussian and anisotropic,
and may modify the short distance dispersion relations.

\end{abstract}
\pacs{PACS numbers:  98.80.Cq,12.90.+b,11.25.w}
}%end tighten
]\narrowtext %% end two-column
%------%%%%%%%%%%%%%%%%%%%%%%%%%%%%%%%%%%%%%%%%%%%%%%%%%%%%%%%%%%%%%%%%

It has long been recognized that cosmology provides a fertile testing
ground for theories beyond the standard model of particle physics.
For string theory, in fact, cosmology may one day  provide the most
accessible way to probe the
theory experimentally. In this regard, inflation is an especially
promising framework as the enormous growth of scales in
the early universe stretches
regions on the order of the Planck scale --- the likely relevant
scale for string theory --- to the much larger
scales of relevance for cosmology.

Recently, there has been significant interest in noncommutative
geometry due  to developments in matrix theory \cite{matrix}
and the realization
\cite{DH,CDS,CH1,volker,sw}
that noncommutative spacetime arises naturally in string and M-theory
when a certain background gauge field is turned on.
In particular, it was shown \cite{CH1,volker,sw}
that in the presence of a constant NS
$B_{\m\n}$-field, the endpoints of the open string obey the
commutation relation
\be \label{comm}
[x^\mu, x^\nu] = i \th^{\mu \nu},
\ee
where the $\th^{\mu \nu}$ are entries of an antisymmetric real
constant matrix of dimension length squared.
The relation between $\th$ and $B$ can be found in \cite{CH1,volker,sw}.
Moreover the commutation relations of the string modes
are modified.
These relations have been employed directly to construct
noncommutative open string theory at any loop order \cite{CRS}.
Note that perturbatively  the noncommutativity is only felt by open strings,
closed strings  are not affected by the $B$-field.

A number of authors have studied the possible phenomenological
effects of such noncommutativity \cite{NCQED2}.
In this brief note, using basic properties of
noncommutative field theory \cite{sw,filk,uvir,wilson}, we point out
some possible cosmological signatures. The idea is that if spacetime
is indeed noncommutative on short distance scales, this may
have an impact on early universe physics. As above, we
work in the context of inflation which allows such short scale
noncommutativity to amplify into large scale cosmological implications.
Specifically, we focus on the generation of density perturbations.
In the usual setup, quantum fluctuations of the inflaton field,
after suitable tuning of its potential, can give rise to the requisite
density perturbations for structure formation. We reexamine these
calculations in the noncommutative framework and point out 
features that can differ from the usual commutative case.
%rr
Inflationary cosmology in noncommutative
geometry from a different perspective was studied in \cite{mangano}.

It is convenient to work in the dual language
 of fields whose algebraic structure is defined by the Moyal product
\be \label{prod1}
(f * g) (x) = e^{i \frac{\th^{\mu\nu}}{2} \frac{\del}{\del \xi^\mu}
\frac{\del}{\del \zeta^\nu} } f(x+ \xi) g(x+\zeta) |_{\xi=\zeta=0},
\ee
which is associative,  noncommutative and satisfies
\be \label{integ}
\int f*g = \int g*f = \int f g .
\ee
Using this $*$-product,
field theory on a noncommutative space (ie. $\th^{\m\n} \neq 0$ only
for $\m,\n \neq 0$)
can be easily formulated. Since it is not clear how to quantize a theory
with
nonzero $\th^{0i}$ \cite{gomis,zamora}, we will restrict ourselves to
spatial
noncommutativity.
Realizing a noncommutative field theory simply amounts to
replacing the usual multiplication of functions by the $*$-product. For
example, noncommutative QED is given by
\be
S=  -\frac{1}{4} \int d^4x (F_{\mu \nu} * F^{\mu \nu}
+ i \bar{\psi} * \Ds *  \psi )
\ee
where $F_{\mu \nu} = \del_\mu A_\nu - \del_\nu A_\mu + i g [A_\mu,
A_\nu]_*$ and
$D_\mu \psi = \del_\mu \psi +i g A_\mu * \psi  $ for a Dirac
spinor.
Note that due to \eq{integ}, the quadratic part of the action
is not modified by $\th$. The theory is non-local as the
$*$-product gives rise to an infinite number of
derivatives in the action.

%XXX
Before we turn to inflation,
we remark that when quantum effects are taken into account,
the naive $\th \rightarrow 0$ limit may not be smooth  in the sense that
in this limit a noncommutative field theory
does not always reduce to its commutative $\th = 0$ counterpart
%a commutative theory in the $\th \rightarrow 0$ limit
\cite{uvir,martin0,ha,cs}.
For example,
the beta function for  noncommutative QED is found to be
\cite{martin0,ha,uvir}
\be
\b = - \frac{g^3}{16 \pi^2} (\frac{22}{3} - \frac{4}{3} N_f)~,
\ee
where $N_f$ is the number of Dirac fermions.
Note that the new (negative) term is due to
the  self-interaction of the noncommutative photons and is
independent of $\theta$ so long as $\theta$ is nonzero.
Summing
together  the contributions from the standard model matter fields,
one finds that the beta function is negative. This is in
conflict with our expectation that the $U(1)$ coupling is not
asymptotically free.
Moreover once $\th$ is turned on,
gauge invariance and
the fact that some standard model fields are charged under both $U(1)$
and $SU(2)\times SU(3)$ imply that the noncommutative gauge symmetry
has to be enlarged to $U(1) \times U(2) \times U(3)$ \cite{sw}.
With $U(3)$ as the color group, the existence of baryons becomes a
problem.
However by
embedding the noncommutative theory in string theory, one may be
able to resolve these issues
with the addition of new degrees
of freedom \cite{uvir} which become effective at the scale
$1/\sqrt{\th}$. More work has to be done to
substantiate this picture.
These new degrees of freedom have implications for the
signatures studied in \cite{NCQED2}.

The above considerations suggest
an alternative appealing framework in which conventional commutative
geometry
emerges from a fundamental noncommutative starting point:
the degree of noncommutativity may be scale-dependent (or
temperature-dependent with $\Lambda$ replaced by $T$ below).
For example, 
\be\label{scaletheta}
\th^{\m\n}=
\cases{\th^{\m\n}, & \mbox{if $\L > \L_{0}$}, \cr
 0 ,          & \mbox{if $\L < \L_{0}$}. }
\ee
We note that \eq{scaletheta} is not the same as
a spatially varying $\th$ and  that
a scale or temperature dependent $\theta$ is consistent
with associativity.
To our knowledge, this simple possibility has not been discussed
before.
The scale $\L_0$ can be interpreted as the Wilsonian cutoff 
of the field theory. 
As long as 
$\L_0$ is much higher than the electroweak  or 
SUSY breaking scale, the problems mentioned above can be avoided.
%to avoid inducing too much instability of
%proton.  
An interesting scenario is to suppose that
$\L_0$ is significantly larger than the electroweak scale, 
but smaller than the
scale of inflation (which is roughly the GUT scale if the inflaton 
is embedded
in a GUT model, or the Planck scale in models of chaotic inflation) 
so that one has a noncommutative universe during
the inflationary period 
%rr
\footnote{Recently, an interesting scenario in which
the commutative world is recovered in the low energy regime, 
together with a decoupling of the above mentioned problematic 
$U(1)$ degree of freedom and with supersymmetry 
broken dynamically is discussed in \cite{CKT}.
}.  
As we now discuss,
since the dynamics of the inflaton field in such a scenario is governed by a
noncommutative field theory which is non-local and Lorentz 
non-invariant, 
the density perturbations due to quantum fluctuations
of the inflaton field are different from that found in usual inflation.

One of the central ideas of modern cosmology is that the observed
inhomogeneity of the universe has its origin in the quantum
fluctuations of fields that are present
during inflation \cite{density,lyth}. These quantum fluctuations,
generated during the ``slow rolling'' period were initially taken
outside
the horizon by the rapid inflation and their form was frozen
until they re-entered the horizon. These
primordial perturbations  then grew with time
due to gravitational instability and  eventually became
the observed classical structures of the universe.

The precise form of these fluctuations depends on the kinematics and
dynamics of the inflaton field. For example, in
the simplest inflationary models, the quantum fluctuations
have a gaussian distribution (for a review of inflationary cosmology,
see for example \cite{KTL}), with amplitudes governed by free
field dynamics. But
non-gaussian perturbations are possible in
more complicated models.
(For example, higher derivative inflationary dynamics was considered in
\cite{AGW}. Interestingly, the interactions in a noncommutative field
theory
generically contain higher derivative terms of the kind in \cite{AGW}.)
Here we note that even in free noncommutative field theory,
the kinematics are such that
a non-gaussian distribution is naturally
obtained. The deviation from  gaussian processes is determined by
the magnitude of the noncommutativity parameter $\theta$.
We also note that the dynamics of noncommutative field theory
can lead to yet other deviations from traditional density perturbation
calculations. 

For simplicity,  we assume that the noncommutativity of the universe
at the inflation scale takes the form of
\eq{prod1}. Since we need
to consider products of fields at different  points, we
also need the more general $*$-product \cite{wilson}
\be \label{prod2}
f (x_1)* g(x_2)  =  e^{i \frac{\th^{\mu\nu}}{2} \frac{\del}{\del x_1}
\frac{\del}{\del x_2} } f(x_1) g(x_2) ,
\ee
%which generalizes
%\eq{prod1} and is defined for functions defined at different points. It
This is easily shown to be associative.
%is easy to show the ``associativity''
%$ f(x_1) * (g (x_2)*h(x_3)) = (f(x_1) *g(x_2) )*h(x_3).$

Now we want to study  quantum fluctuations of the inflaton $\phi$.
During the slow roll period of 
inflation, the potential
energy $V(\phi)$ is approximately constant and the universe can be
described by the de Sitter spacetime
\be
ds^2 = dt^2 - e^{2Ht} d\bx^2.
\ee
We will assume that the inflaton field comes from
the open string sector, even though {\em a priori}, it can be
a closed string state. However, this assumption
is quite natural in brane-world inflationary scenarios, (see {\em e.g.},
\cite{dvali-tye}).
The action for the inflaton $\phi$ is
\be \label{S0}
S =\int d^4 x \sqrt{-g} \left(\frac{1}{2} g^{\m\n} \del_\m \phi *
\del_\n\phi - V_*(\phi) \right).
\ee
Here we take gravity as a background that is
not affected by the noncommutativity.
More general considerations with
gravity also seeing the noncommutativity can be found in
\cite{grav}.
Since $\th^{0i}=0$ and the metric is independent of $\bx$,
\eq{integ} can be generalized with an arbitrary time-dependent factor
inserted.
%rr
Note that the $*$-product in \eq{S0} is taken to be defined with
respected to the comoving coordinates $\bx$. That is to say we 
are considering a sceranio in which noncommutativity parameter
$\th^{\m\n}$ is constant in the comoving frame.
In physical coordinates
this means that $\theta$ is growing with the scale factor, something 
that seems reasonable since the B-field is the superpartner of the
metric. If $\theta$ subsequently drops to zero, say at the end of inflation,
this should yield a viable cosmology
\footnote{ A different scenario
in which $\th^{\m\n}$ is constant in the physical coordinates can also
be considered. The difference between the two is the cosmological scaling
factor. This will have an important difference in 
the observational predications of the model. We will comment on this
later.}.  
One obtains
the equation of motion
\be \label{eom1}
\ddot{\phi} + 3H \dot{\phi} - e^{-2 H t} \Delta \phi + {\d V_*}/{\d
\phi} =0,
\ee
where $\Delta$ is the usual 3-dimensional Laplacian.
Until a few Hubble times after the horizon exit,
one can drop the $V_*''$ term \cite{lyth}.
In more complicated models,
effects of the potential will have to be taken into account. We
will take $\phi$ to be free, except for
some general comments at the end. Even in this case, we will show
that noncommutativity can yield deviations from the usual
gaussian density perturbations.
%rr
This is purely a result of the ``background''
noncommutative geometry. 
The equation for the
fluctuations thus takes the free form
\be
\ddot{\phi} + 3H \dot{\phi} - e^{-2 H t} \Delta \phi =0.
\ee
Here $\phi$ represents
the fluctuating part of the inflaton;  it has the Fourier expansion 
($ k = |\bk|$),
\be
\phi(\bx,t) = \int_\bk
\frac{1}{\sqrt{2 k}}
(a_\bk \psi_\bk(t) e^{i \bk \cdot \bx} +
h.c.),
\ee 
%rr
where $\bk$ is the wave vector in the comoving frame, 
$
\psi_\bk(t) = \frac{i H}{k} ( 1+ \frac{k}{i H} e^{-H t} )
\exp( \frac{i k}{H} e^{-H t}) ,
$
$ \int_\bk \equiv \int \frac{d^3\bk}{(2\pi)^3}$
and $ \psi_\bk(t) \sim e^{- ik t}
\quad\mbox{for} \quad k/H \gg 1. $
Canonical quantization of $\phi$ imposes the commutation relations
\be
[a_\bk, a^\dagger_{\bk'}] = (2\pi)^3 \d(\bk-\bk'),
\quad [a_\bk, a_{\bk'}] =0.
\ee

Since $\phi$ feels the noncommutativity,
the relevant $n$-point correlation function is
\be
I_{n}(\bx_1, \cdots, \bx_n) =
\langle 0| \phi(\bx_1,t) * \cdots* \phi(\bx_n,t) |0 \rangle,
\ee
where the time dependence is understood.
Essentially as in \cite{filk} one obtains
\be
I_2 = \int_\bk \frac{e^{i \bk \cdot(\bx_1-\bx_2) } }{2k}
\eta_k (t), \quad \eta_k (t) \equiv e^{-2Ht} + {H^2}/{k^2}
\ee
which is independent of $\th$ and takes the usual form.
As for the 4-point function, it is
\bea \label{4point}
I_4 &&=I_2(\bx_1-\bx_2) I_2(\bx_3-\bx_4)  +
I_2(\bx_1-\bx_4) I_2(\bx_2-\bx_3) + \nn\\
+ &&\int_\bk \int_{\bk'}
\frac{e^{i \bk \cdot(\bx_1-\bx_3) } }{2k} \eta_k (t)
\frac{e^{i \bk' \cdot(\bx_2-\bx_4) } }{2k'} \eta_{k'} (t)
e^{-i \bk \wedge \bk'},
\eea
where
$ k \wedge k' \equiv k_\m \th^{\m\n} k'_\n $, with $I_3 = 0$ 
and more generally  $I_{2n+1}=0$.
Note that although $\phi$ is
a free field, the $n$-point functions depend on $\th$ because
of the $*$-product.  Note also that because of these contributions,
even in free field theory
$I_4$ and generally $I_n$ cannot be factorized in terms of products of
$I_2$.
A few Hubble times after horizon exit,
the 4-point function in momentum space is
\bea\label{4pt-mom}
&&I_4(\bk_1,\bk_2,\bk_3,\bk_4)=(4\pi^3 H^2 k_1^{-3})\cdot \nn\\
&&\;\;[ k_3^{-3}\d(\bk_1+\bk_2)\d(\bk_3+\bk_4)
+ k_2^{-3}\d(\bk_1+\bk_4)\d(\bk_2+\bk_3) \nn\\
&& \;\;+k_2^{-3}\d(\bk_1+\bk_3)\d(\bk_2+\bk_4)e^{-i \bk_1 \wedge
\bk_2}].
\eea
These quantum fluctuations are carried outside the horizon
and lead to curvature fluctuations.
By the time
these fluctuations re-enter the horizon, 
the relevant physical processes occur at
a lower scale and hence are described by a commutative dynamics.
%cb since the temperature has dropped below $T_0$.
Therefore, they will just set the
initial conditions for the subsequent evolution of the
perturbations
and show up as the observed inhomogeneities of the universe;
see for example \cite{mfb} for a comprehensive treatment.
Since $I_2$ is unmodified, we get
the usual power spectrum $P(k) = k^{n-1}$, with spectral index $n=1$ for
the free case.
However, since $I_4$ does not factorize into products of
$I_2$, the subsequent
distribution will not be gaussian. We note that this violation of
 gaussian statistics is independent of the couplings and is universal;
in any specific scenario there may be additional model-dependent violations.
The non-gaussian fluctuations
will be reflected, for example, in the galaxy distribution and the CMB
measurements;
stronger constraints are expected to come from the latter.
By extracting the 4-point function from the existing 4-year DMR maps
or more refined sky maps from future experiments such as MAP and PLANCK,
one should be able  to set a 
bound on the degree of noncommutativity during inflation.
%one can against \eq{4pt-mom} and should be able to
%set a bound on the degree of noncommutativity during inflation.

%rr  
% add paragraph on estimation of theta 
In the above, we considered the case in which $\th$ is constant
in the comoving frame. As we mentioned, 
one may also consider the case in which $\th$
is constant in the physical frame. All of the above
formula are basically unchanged, except that we just have to use 
a time varying $\th^{\m\n}(t)$,
\be
\th^{\m\n}(t) = e^{-2H t} \th_0^{\m\n}.
\ee  
Due to the exponential decay factor, 
the effect of noncommutativity gets redshifted away by inflation
\footnote{We thank Will Kinney for important discussions on this point.}.
However so long as $\th$ is constant
in the comoving frame and shuts down by the end of inflation, 
this can lead to a small amount of non-gaussianity.
Again, we remark that 
a comoving constant $\th$ is suggested in
string theory since the NSNS $B$-field and the metric $g$ are 
coupled to each other through their equation of motion. It is thus natural
to assume that $\th$ grows since the metric may grow by inflation.

%%%%

Beyond the universal effects due to 
noncommutative geometry discussed so far, there will be
additional effects arising from the dynamical
details of any particular model. One expects that
noncommutative interactions will make a
$\theta$ dependent contribution to the spectrum of fluctuations
(similar to the analysis in \cite{GMS}).
In the commutative case, interaction effects 
are often too small to be observed,
so  it is worth determining if
there are noncommutative models in which their impact
is amplified.
We also note that nonzero $\theta$ may potentially be relevant for
understanding the CMB dipole anisotropy.
The CMB dipole from DMR has an amplitude $3.358\pm0.024$ mK
in a particular known direction\cite{smoot}.
The conventional
interpretation invokes
the Doppler effect arising from the motion of
the solar system with respect to the CMB rest frame. There
is room, however, for other possible contributions to the dipole
anisotropy.
For instance, nonzero
$\theta$ introduces a degree of anisotropy whose contribution will
depend in detail on the form of the interactions coded in $V_*$.
Whether this yields a viable contribution to the dipole is thus
a model dependent question that we leave to future work.

Frameworks for studying related issues have been developed in
\cite{mb} and \cite{kempf}, in which modifications to conventional physics
at sub-Planck scales are modeled and their effects on inflation
are studied. In \cite{kempf}, the focus is on a quantized
spacetime\cite{qs} and the string uncertainty relations \cite{sur}. 
In \cite{mb}, the authors study the effects on inflation due to
so called  {\it trans-Planckian} dispersion relations, which have
been postulated \cite{unruh,jacobson}  to encode  strong  gravity
effects at sub-Planck scales.
We note that even in the absence of gravity, 
the dispersion relations of a noncommutative quantum field theory
are also  modified  by loop effects \cite{uvir,susskind,espe}.  
Since gravity will generally not just
modify the propagator, 
but will also introduce new interactions into the theory, the
trans-Planckian dispersion relations are expected to be further corrected.
One should then study the
combined effects of both on the short distance dispersion relations and
determine the impact on the primordial spectrum of perturbations.
We leave this interesting analysis for future work.

%%%%%%%%%%%%%%%%%%%%%%%%%%%%%%%%%%%%%%%%%%%%%%%%%%%%%%%%%%%%%%%%%

In this letter we focused mainly on the case of a scale dependent
$\th$. 
A related scenario is that the world is
commutative at low temperature but becomes more and
more noncommutative (and non-local)
once the temperature is higher than a certain threshold temperature
$T_0$.
In string theory, noncommutativity arises when a non-zero background
$B$-field is turned on. In perturbative string theory, this
$B$-field is a modulus, and so its value
is arbitrary. However, in four dimensions,
the NS $B$-field is dual to a scalar. And
just like the dilaton, it is possible that
a potential can be generated for $B$ non-perturbatively.
For example, if an
isotropic potential of the following inverse
symmetry breaking form (see e.g. \cite{Weinberg})
\begin{equation}
V(X) = {\mu \over 2} ~\left[ 1- \left({T \over T_0}\right)^2 \right] X^2
+ {\lambda \over 4} X^4
\end{equation}
is generated for $X^2=\bf{B}^2$ and $\mu$, $\lambda>0$,
then for $T<T_0$, the minimum of the potential is
at $X=0$,  and spacetime is commutative. For $T>T_0$, $X =0$ is a maximum
and the true minimum occurs at
$
X^2 ={{\mu \over \lambda}
((T/ T_0)^2 -1) } ~.
$
As a result, spacetime becomes noncommutative and
rotational invariance is broken at temperature $T>T_0$.
Moreover the degree of noncommutativity depends on $T$. 
A more thorough understanding of how this model is embedded in string theory
and  how Lorentz symmetry is restored in the low energy limit 
would be desirable. 
We expect that a temperature dependent $\th$ will have interesting
consequences on the thermal history of a hot big bang universe.

While we do not pursue it in this paper,
our setup can be
embedded naturally in the brane world scenario
\cite{add,ST,BW,ovrut,RS,kir},
if our
four-dimensional world is localized on a brane whose worldvolume
has a non-zero background $B$-field in the early universe.
The cosmological implications of this scenario have been studied
in some detail
(see, {\it e.g.},
\cite{dvali-tye,branecosmology}).
Here we expect additional new features as the universe undergoes
a period of noncommutativity.

Finally, we remark that
noncommutativity may also appear in the extra
dimensions \cite{lust,wise} in which case $\th$ can be taken to
be scale independent.
In the brane world  scenario,
this extra-dimensional  noncommutativity
arises when a higher dimensional brane is wrapped
around the extra compactified
directions in the presence of a constant $B$-field.
With enough supersymmetry, the universe
is effectively four-dimensional and commutative at
energies below the threshold of the Kaluza-Klein modes, and
$\th$-modifications are possible only through loops.
At energies higher than the Kaluza-Klein threshold, noncommutativity
becomes
effective. This
clearly has implications for
collider experiments as well as for the early universe.
In the latter case,
one has to study the implications for the quantum fluctuations of the
higher dimensional noncommutative inflaton.
Many interesting questions about this scenario await
to be explored.
We plan to address some of these issues
in future works.

\vspace{.5cm}

The work of C.S.C. was supported by the Swiss National Science
Foundation and by the European Union under RTN contract HPRN-CT-2000-00131.
The work of B.R.G. was supported by DOE grant DE-FG02-92ER40699B.
The work of G.S. was supported in part by
the DOE grant FG02-95ER40893 and the University of Pennsylvania
School of Arts and Sciences Dean's fund.
We thank Adel Bilal, Robert Brandenberger, Jean-Pierre Derendinger,
Chung-Pei Ma, Kostas Sfetsos, Henry Tye and especially Will Kinney
for useful discussions.

\ed